\documentclass[traditabstract]{aa}

\usepackage{natbib}
\usepackage{amssymb,verbatim}
\usepackage{graphicx}
\usepackage{hyperref}

\begin{document}

\title{The stripping of a galaxy group diving into the massive cluster A2142}

\author{D. Eckert\inst{1,2} \and S. Molendi\inst{2} \and M. Owers\inst{3} \and M. Gaspari\inst{4,5} \and T. Venturi\inst{6} \and L. Rudnick\inst{7} \and S. Ettori\inst{5,8} \and S. Paltani\inst{1} \and F. Gastaldello\inst{2,9} \and M. Rossetti\inst{10,2}}
\institute{
Department of Astronomy, University of Geneva, ch. d'Ecogia 16, 1290 Versoix, Switzerland\\
\email{Dominique.Eckert@unige.ch}
\and
INAF - IASF-Milano, Via E. Bassini 15, 20133 Milano, Italy
\and
Australian Astronomical Observatory, PO Box 915, North Ryde, NSW 1670, Australia
\and
Max Planck Institute for Astrophysics, Karl-Schwarzschild-Strasse 1, 85741 Garching, Germany
\and
INAF - Osservatorio Astronomico di Bologna, Via Ranzani 1, 40127 Bologna, Italy
\and
INAF - Istituto di Radioastronomia, via Gobetti 101, 40129, Bologna, Italy
\and
Minnesota Institute for Astrophysics, School of Physics and Astronomy, University of Minnesota, 116 Church Street SE, Minneapolis, MN 55455, USA
\and
INFN, Sezione di Bologna, viale Berti Pichat 6/2, 40127 Bologna, Italy
\and
Department of Physics and Astronomy, University of California at Irvine, 4129 Frederick Reines Hall, Irvine, CA 92697-4575, USA
\and
Universit\` a degli studi di Milano, Dip. di Fisica, via Celoria 16, 20133 Milano, Italy
}

\abstract{Structure formation in the current Universe operates through the accretion of group-scale systems onto massive clusters. The detection and study of such accreting systems is crucial to understand the build-up of the most massive virialized structures we see today. We report the discovery with \emph{XMM-Newton} of an irregular X-ray substructure in the outskirts of the massive galaxy cluster Abell 2142. The tip of the X-ray emission coincides with a concentration of galaxies. The bulk of the X-ray emission of this substructure appears to be lagging behind the galaxies and extends over a projected scale of at least 800 kpc. The temperature of the gas in this region is 1.4 keV, which is a factor of $\sim4$ lower than the surrounding medium and is typical of the virialized plasma of a galaxy group with a mass of a few $10^{13}M_\odot$. For this reason, we interpret this structure as a galaxy group in the process of being accreted onto the main dark-matter halo. The X-ray structure trailing behind the group is due to gas stripped from its original dark-matter halo as it moves through the intracluster medium (ICM). This is the longest X-ray trail reported to date. For an infall velocity of $\sim$1,200 km s$^{-1}$ we estimate that the stripped gas has been surviving in the presence of the hot ICM for at least 600 Myr, which exceeds the Spitzer conduction timescale in the medium by a factor of $\gtrsim400$. Such a strong suppression of conductivity is likely related to a tangled magnetic field with small coherence length and to plasma microinstabilities. The long survival time of the low-entropy intragroup medium suggests that the infalling material can eventually settle within the core of the main cluster.}

\maketitle

\section{Introduction}

Galaxy clusters are thought to have grown through the continuous accretion and merging of smaller halos across cosmic time \citep[e.g.][]{springel06}. These processes are expected to be still active in the outer regions of massive local clusters, through the accretion of galaxies and groups of galaxies onto the main dark-matter halo. Accreting substructures in the form of infalling galaxies and galaxy groups contribute significantly to the growth of galaxy clusters in mass, member galaxies and hot gas \citep[e.g.][]{berrier09,genel10,delucia12}. Therefore, such events provide us with excellent laboratories for the study of the processes leading to the assembly of galaxy clusters.

There are many examples of substructures observed in the optical \citep[e.g.][and references therein]{girardi02}, indicating that the majority of clusters are still dynamically active. The dynamics of small structures is however complicated to infer directly from optical data because of the small number of members, and their association with an underlying dark-matter halo is difficult to assess \citep{lemze13}. Weak-lensing studies \citep[e.g. LoCuSS,][]{okabe10} are effective at detecting massive halos, but are insufficiently sensitive to minor mergers (i.e. mergers with mass ratio $\sim$1/100 or lower). The detection of extended X-ray emission provides a cleaner way of assessing the presence of low-mass infalling halos, since hot gas is an unambiguous signature of a virialized halo. 

While X-ray observatories like \emph{Chandra} have provided spectacular examples of merging clusters \citep[and references therein]{mark07,owers09}, the accretion of smaller halos onto massive clusters has proven more elusive. During infall, the ram pressure applied by the ambient ICM is responsible for stripping the gas from its original halo and heating it up \citep{gunn72,vollmer01,heinz03,roediger08}, leading to the virialization of the gas in the main dark-matter halo. This process has been observed in a handful of interacting galaxies in the nearest clusters like Virgo \citep[e.g.][]{machacek04,randall08}, A3627 \citep{sun07,zhang13}, Pavo \citep{machacek05} and A1367 \citep{sun05}. 

In the local Universe, groups with a gravitational mass in the range of a few $10^{13}M_\odot$ are the largest reservoir of cosmic gas and total matter \citep{fukugita98}. Thus, at the present epoch the continuous infall of such groups is the major process by which clusters grow in the current favored scenario of hierarchical structure formation in a cold dark matter dominated Universe. The $\Lambda$CDM paradigm predicts that there should be approximately $\sim$1 such accreting group per massive cluster and per Gyr \citep{dolag09}. There are a few examples of accreting groups directly observed in X-rays such as NGC 4839 in Coma \citep{briel92,neumann01} and the southern group in A85 \citep{kempner02,durret05}. However in both instances, the bulk of the X-ray emission is localized around the dominant elliptical galaxy of the group.

Abell 2142 is a massive \citep[$M_{200}=1.3\times10^{15}M_\odot$,][]{munari13}, X-ray luminous galaxy cluster at a redshift of 0.09 \citep{owers11}. This cluster is known to be dynamically active, as highlighted by its X-ray morphology elongated in the NW-SE direction \citep{buote96}, the presence of prominent cold fronts \citep{mark00,rossetti13}, and of Mpc-scale radio emission \citep{farnsworth13}. A dynamical analysis of $\sim$1,000 member galaxies indicates the presence of several small subgroups \citep{owers11}. 

We obtained an \emph{XMM-Newton} mosaic of this cluster covering the entire azimuth out to the virial radius. In this paper, we report the discovery in this program of an accreting galaxy group in the outskirts of A2142. The paper is organized as follows. In Sect. \ref{sec:data}, we introduce the available data and the analysis procedure. In Sect. \ref{sec:results} we present our results and the existing multiwavelength information (optical and radio). The implications of our findings are discussed in Sect. \ref{sec:disc}.

Throughout the paper, we assume a $\Lambda$CDM cosmology with $H_o=70$ km s$^{-1}$, $\Omega_m=0.3$ and $\Omega_\Lambda=0.7$. At the redshift of A2142 ($z=0.09$), this corresponds to $1^{\prime\prime}=1.7$ kpc. For an average cluster temperature of 9.0 keV \citep{sabrina02} and using the scaling relations of \citet{arnaud05}, we estimate $R_{500}=1450$ kpc $=14.4$ arcmin\footnote{For a given overdensity $\Delta$, $R_\Delta$ is the radius for which $M_\Delta/(4/3\pi R_\Delta^3)=\Delta\rho_c$} and $R_{200}=2200$ kpc $=21.6$ arcmin. All the quoted error bars are at the $1\sigma$ level.

\section{Data analysis}
\label{sec:data}
\subsection{Imaging}
\label{sec:imaging}
Abell 2142 was mapped by \emph{XMM-Newton} in 2011 and 2012 through a central 50 ks pointing \citep[OBSID 067456, see][]{rossetti13} and four 25 ks offset pointings (OBSID 069444), which allowed us to obtain a coverage of the entire azimuth of the cluster out to $R_{200}$. The data were processed using the \emph{XMM-Newton} Scientific Analysis System (XMMSAS) v13.0. We extracted raw images in the [0.7-1.2] keV band for all three EPIC detectors (MOS1, MOS2 and pn) using the Extended Source Analysis Software package \citep[ESAS,][]{xmmcat}. The choice of the [0.7-1.2] keV energy band is motivated by the fact that this band maximizes the source-to-background ratio \citep{ettoriwfxt} and avoids the bright Al and Si background emission lines, whilst maintaining a large effective area since the collecting power of the \emph{XMM-Newton} telescopes peaks at 1 keV. Exposure maps for each instrument were created, taking into account the variations of the vignetting across the field of view. A model image of the non X-ray background (NXB) was computed using a collection of closed-filter observations and was adjusted to each individual observation by comparing the count rates in the unexposed corners of the EPIC detectors. Point sources were detected down to a fixed flux threshold and excluded using the ESAS task \texttt{cheese}.

\begin{figure}
\centerline{\resizebox{\hsize}{!}{\includegraphics{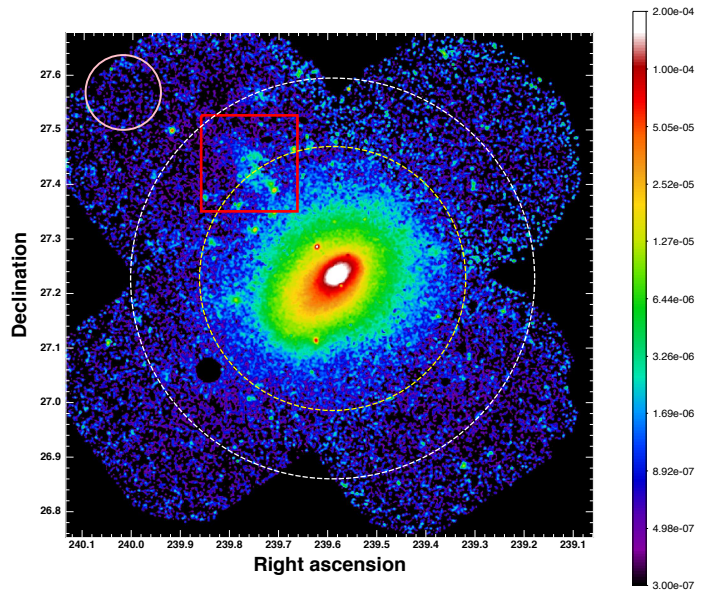}}}
\caption{Exposure-corrected, NXB-subtracted combined \emph{XMM-Newton}/EPIC mosaic image of A2142 in the [0.7-1.2] keV band, smoothed with a Gaussian of width 15 arcsec. The units for the color bar are MOS counts s$^{-1}$. The dashed circles show the approximate locations of $R_{500}$ and $R_{200}$, while the red square indicates the position of a newly-discovered X-ray substructure in this cluster. The pink circle in the top-left corner shows the region used to estimate the local background components.}
\label{fig:mosaic}
\end{figure}

We computed surface-brightness images by subtracting the NXB from the raw images and dividing them by the exposure maps. To maximize the signal-to-noise ratio, we then combined the surface-brightness images of the three EPIC detectors by weighing each detector by its relative effective area, and created a combined mosaic image of the cluster using all 5 observations. The total \emph{XMM-Newton}/EPIC image of A2142 is shown in Fig. \ref{fig:mosaic}. The dashed circles indicate the approximate locations of $R_{500}$ and $R_{200}$. An intriguing diffuse, irregular structure is observed at $\sim R_{500}$ northeast (NE) of the cluster core, highlighted by the red square in Fig. \ref{fig:mosaic}. For the remainder of the paper, we focus on this particular structure; the global mosaic will be the subject of a forthcoming paper.

\subsection{Spectral analysis}
\label{sec:spec}

\begin{figure}
\centerline{\resizebox{\hsize}{!}{\includegraphics[angle=270]{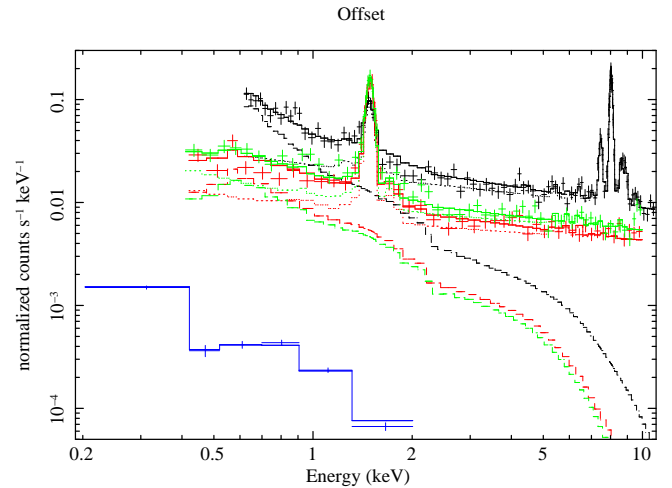}}}
\caption{Spectrum of an offset region located $\sim32$ arcmin north of the cluster core to measure the sky background components, from EPIC pn (black), MOS1 (green), and MOS2 (red), and from \emph{ROSAT}/PSPC (blue). The dotted lines show the model NXB for each instrument, while the dashed lines show the best-fit three-component sky background model.}
\label{fig:skybkg}
\end{figure}

We performed a spectral analysis of this structure using our 25 ks observation of the NE region in A2142. Spectra and response files for each region considered were extracted using the ESAS tasks \texttt{mos-spectra} and \texttt{pn-spectra} and were fit in XSPEC v12.7.1. The spectra were grouped to ensure a minimum of 20 counts per spectral channel. Since the surface brightness of the emission in the NE excess region barely exceeds the background level, a detailed modeling of all the various background components is necessary to obtain reliable measurements of the relevant parameters. We adopted the following approach to model the different spectral components:

\begin{itemize}
\item \textit{The source:} We modeled the diffuse emission in each region using the thin-plasma emission code APEC \citep{apec}, leaving the temperature, metal abundance and normalization as free parameters. This component is absorbed by the Galactic hydrogen column density along the line of sight, which we fix to the 21cm value \citep[$N_H=3.8\times10^{20}$ cm$^{-2}$,][]{kalberla}.
\item \textit{The non X-ray background (NXB):} We used closed-filter observations to estimate the spectrum of the NXB component in each region, following the procedure described in \citet{xmmcat}. Instead of subtracting the NXB component, we approximated it by a phenomenological model, which we then included as an additive component in the spectral modeling \citep{lm08}. This method has the advantage of retaining the statistical properties of the original spectrum, which allows the use of C-statistic \citep{cash79}. We left the normalization of the NXB component free to vary during the fitting procedure, which allows for possible systematic variations of the NXB level. The normalization of the prominent background lines is also left free. Since the observation was very weakly contaminated by soft proton flares \citep[IN over OUT ratio of 1.05,][]{lm08}, we do not need to include a component to model the residual soft protons.
\item \textit{The sky background components:} We used an offset region located $\sim31$ arcmin away from the cluster core (see the pink circle in Fig. \ref{fig:mosaic}), where no cluster emission is detected, to measure the sky background components in the region of A2142. We complemented these data with the \emph{ROSAT} all-sky survey spectrum of the same region \citep{backtool}, and fitted the \emph{XMM-Newton} and \emph{ROSAT} data jointly. We modeled the sky background using a three-component model: \emph{i)} a power law with photon index fixed to 1.46 to model the cosmic X-ray background \citep[CXB,][]{deluca04}; \emph{ii)} a thermal component at a temperature of 0.22 keV to estimate the Galactic halo emission \citep{mccammon02}; \emph{iii)} an unabsorbed thermal component at 0.11 keV for the local hot bubble \citep{mccammon02}. The best-fit spectrum for the offset region is shown in Fig. \ref{fig:skybkg}, and the corresponding normalizations for the various components are provided in Table \ref{tab:skybkg}. The best-fit CXB component has a flux of $(2.07\pm0.15)\times10^{-11}$ ergs cm$^{-2}$ s$^{-1}$ deg$^{-2}$ in the 2-10 keV band, which agrees with the value of \citet{deluca04}. To model the sky background in a different region, the normalization of each component was rescaled by the ratio of the corresponding areas, accounting for CCD gaps and bad pixels. The CXB component was left free to vary by $\pm15\%$ to allow for cosmic variance \citep{moretti03}.
\end{itemize}

\begin{table}
\caption{\label{tab:skybkg}Best-fit parameters for the sky background components from Fig. \ref{fig:skybkg}.}
\begin{center}
\begin{tabular}{lc}
\hline
Component & Norm arcmin$^{-2}$\\
\hline
\hline
Local Bubble & $(8.1\pm0.5)\times10^{-7}$ \\
Galactic Halo & $(1.15\pm0.08)\times10^{-6}$ \\
CXB & $(9.7\pm0.7)\times10^{-7}$ \\
\hline
\end{tabular}
\end{center}
\textbf{Column description:} 1: Background component (see Sect. \ref{sec:spec}). 2: Normalization of the model per arcmin$^2$. For APEC (local bubble and galactic halo), this corresponds to $10^{-14}/(4\pi d_A^2(1+z)^2)\int n_en_H\,dV$; for a power law (CXB) this is in units of photons keV$^{-1}$ cm$^{-2}$ s$^{-1}$ at 1 keV.
\end{table}

\section{Results}
\label{sec:results}

\subsection{Morphology of the NE feature}

\begin{figure*}
\centerline{\resizebox{\hsize}{!}{\hbox{\includegraphics{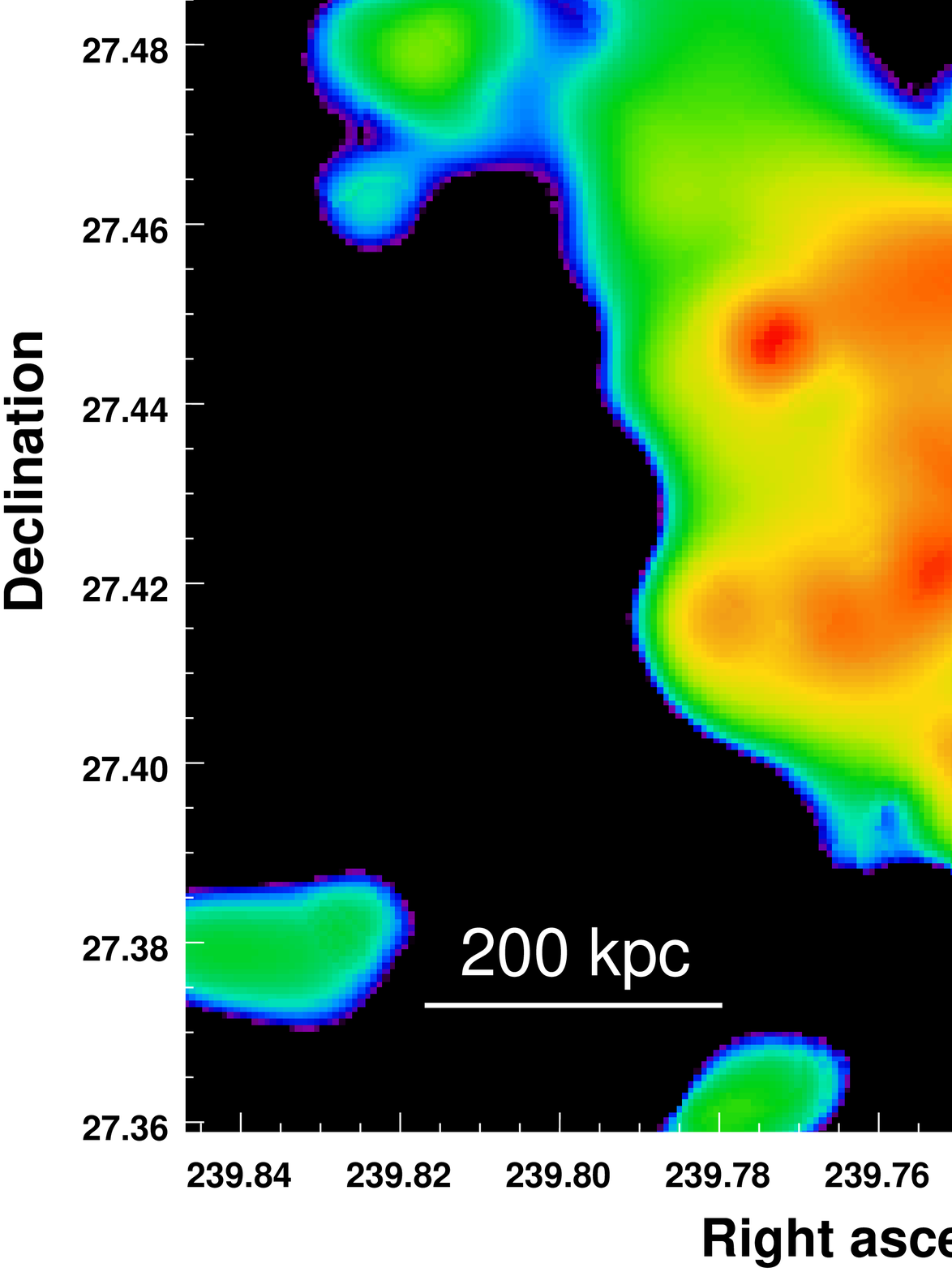}\includegraphics{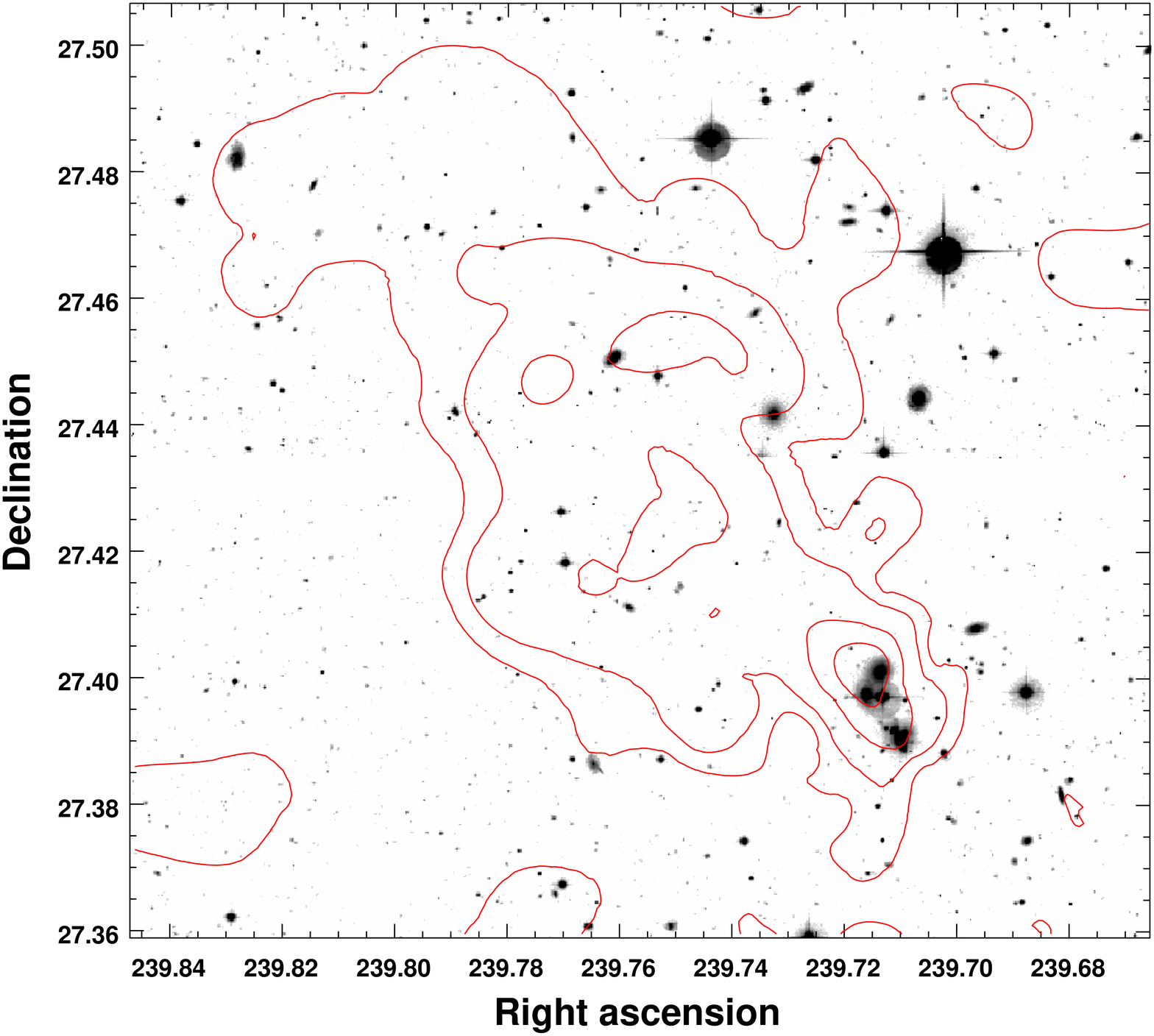}}}}
\caption{\label{fig:zoom}\emph{Left:} Adaptively-smoothed residual image of the NE feature highlighted by the red square in Fig. \ref{fig:mosaic}. \emph{Right:} CFHT/MegaCam $g$-band image of the same region, with X-ray contours overlayed in red.}
\end{figure*}

We studied the morphology of the newly-discovered NE feature using the method described in Sect. \ref{sec:imaging}. To highlight the presence of features in the X-ray image, we created a residual image by subtracting the azimuthally-averaged cluster emission from the EPIC [0.7-1.2] keV mosaic image shown in Fig. \ref{fig:mosaic}. For this purpose, we extracted the surface-brightness profile of the cluster centered on the surface-brightness peak, using the method described in \citet{ccbias1}. We then subtracted the cluster emission from each pixel and masked the detected sources. Finally, the resulting image was adaptively smoothed using the SAS task \texttt{asmooth}. 

In the left panel of Fig. \ref{fig:zoom} we show the resulting residual image zoomed on the NE feature. The feature has an irregular morphology elongated in the northeast-southwest (SW) direction, which ends in a bright ``tip'' at its SW pointing towards the cluster center. The tip hosts an X-ray point source (which has been masked in Fig. \ref{fig:zoom}) surrounded by a region of diffuse emission. In the right panel of Fig. \ref{fig:zoom} we show an archival CFHT/MegaCam $g$-band image of the same region with X-ray contours overlayed. The tip corresponds with a concentration of at least 5 galaxies; this suggests a possible association of the X-ray feature with these galaxies. Apart from this structure, no particular optical substructure is seen in this region of the cluster \citep[see Fig. 12 of][]{owers11}, and no background concentration of galaxies is seen. 

\begin{figure}
\centerline{\resizebox{\hsize}{!}{\includegraphics{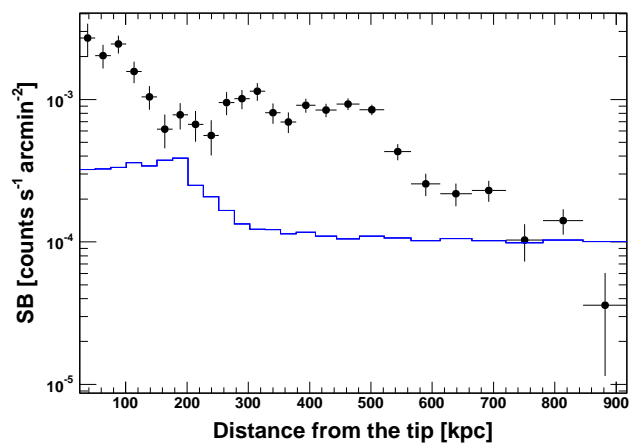}}}
\caption{Background-subtracted EPIC surface-brightness profile in elliptical annuli starting from the tip in the NE direction (position angles 105-165). The sky background intensity was estimated in the pink circle in Fig. \ref{fig:mosaic}. The NXB level is shown by the blue line.}
\label{fig:sbtail}
\end{figure}

Starting from the tip, the X-ray substructure extends in the NE direction over a scale of $\sim8$ arcmin and a width of $2-3$ arcmin. At the redshift of the cluster this corresponds to a projected linear scale of about 800 kpc. We extracted a surface-brightness profile starting from the tip to study the distribution of X-ray brightness. We used elliptical annuli in the sector with position angles 105-165$^\circ$ to follow the morphology of the feature as closely as possible. The NXB was subtracted using the modeling described in Sect. \ref{sec:imaging}. We used the offset region defined in Fig. \ref{fig:mosaic} (see also Fig. \ref{fig:skybkg}) to estimate the local sky background contribution and subtract it from the surface-brightness profile. The resulting profile is shown in Fig. \ref{fig:sbtail}. The surface brightness of the feature is flat out to $\sim500$ kpc, decreasing only by a factor of $\sim3$ compared to the maximum, then decreases more steeply out to 800 kpc, which is the maximum radius at which we are detecting the feature.

\subsection{Spectral properties}
\label{sec:spectra}

\begin{figure}
\centerline{\resizebox{\hsize}{!}{\includegraphics{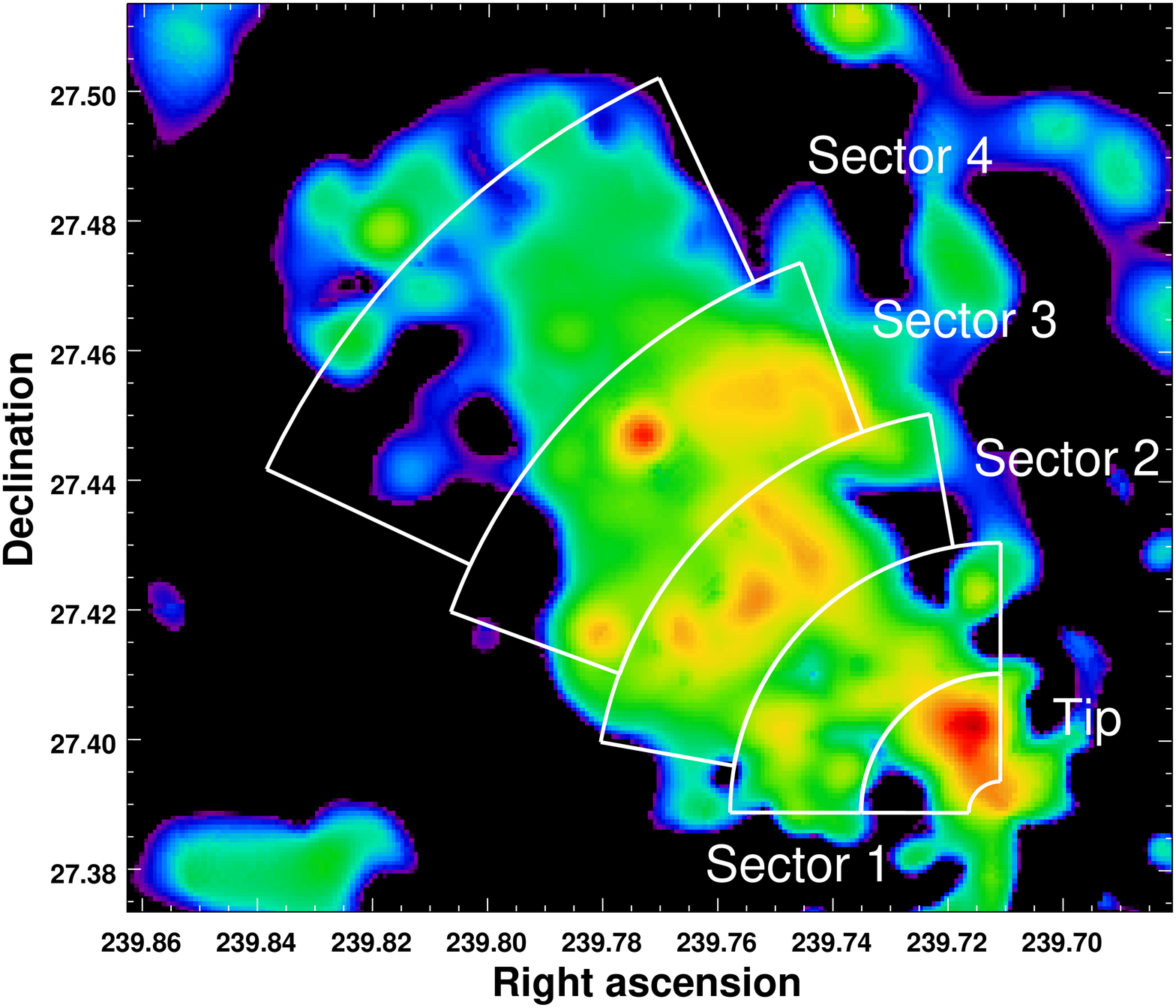}}}
\caption{Regions defined for the spectral analysis.}
\label{fig:regions}
\end{figure}

\begin{figure*}
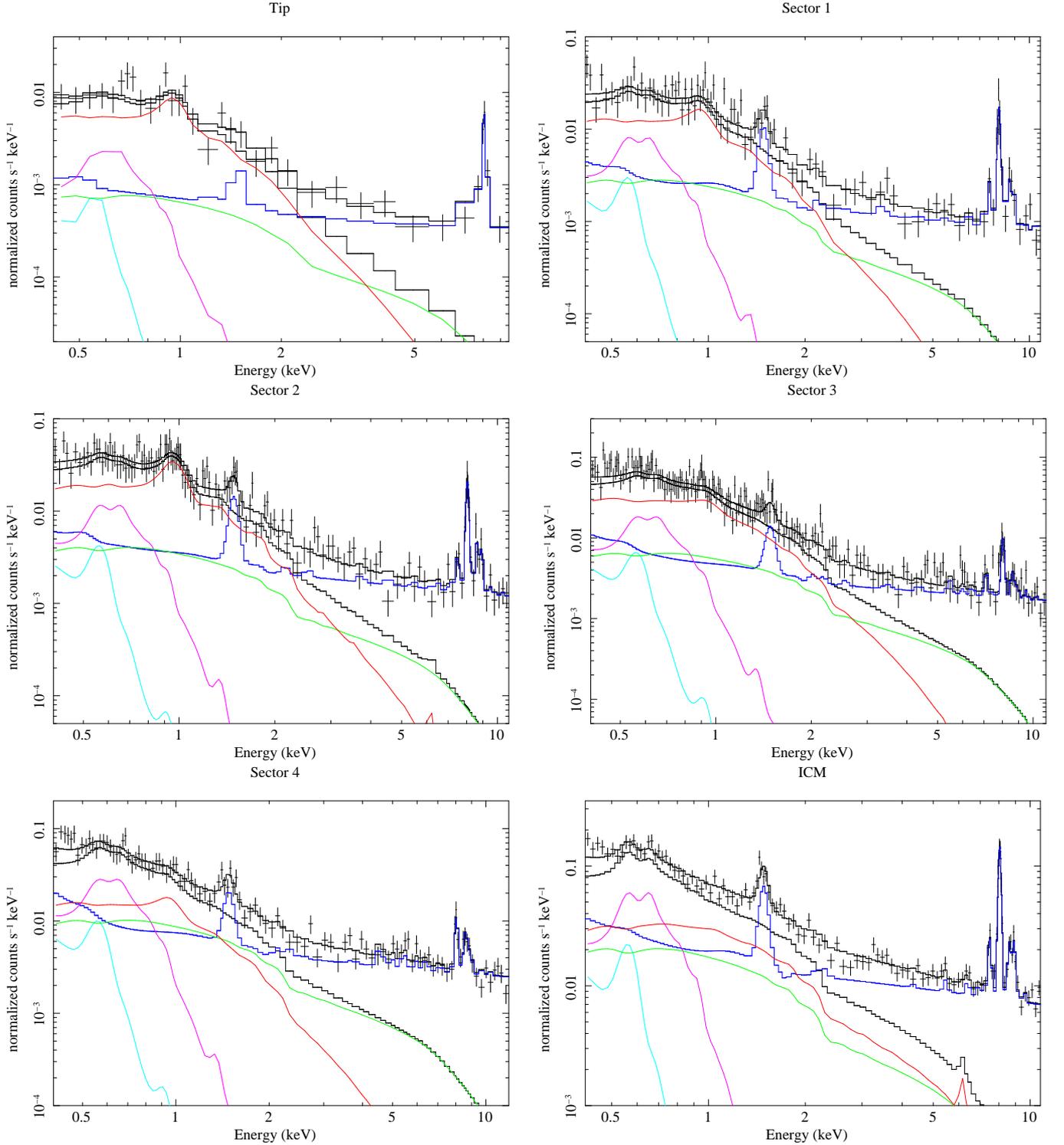

\resizebox{\hsize}{!}{\vbox{\hbox{
\includegraphics[angle=270]{reg1.ps}
\includegraphics[angle=270]{sector1.ps}}
\hbox{
\includegraphics[angle=270]{sector2.ps}
\includegraphics[angle=270]{sector3.ps}}
\hbox{
\includegraphics[angle=270]{sector4.ps}
\includegraphics[angle=270]{icm.ps}}}}
\caption{\emph{First and second rows:} EPIC/pn spectra for the regions defined in Fig. \ref{fig:regions} (the fits were performed jointly on all 3 EPIC instruments, however only pn is shown here for clarity). The solid lines show the various components used for the fitting procedure: the source (red), the NXB (blue), the CXB (green), the Galactic halo emission (magenta), and the local hot bubble (cyan). \emph{Bottom right:} EPIC/pn spectrum of the local ICM in the region close to the tip of the infalling group. The color code is the same as above.}
\label{fig:allspectra}
\end{figure*}

To study the thermodynamical properties of the structure, we extracted the spectrum of the source in 5 sectors along the NE feature. The regions used for the spectral analysis are shown in Fig. \ref{fig:regions}. The ``tip" region was defined excluding the X-ray point source observed there. The following sectors were chosen to trace the morphology of the feature as closely as possible. The spectra and best-fit models for the regions defined in Fig. \ref{fig:regions} are shown in Fig. \ref{fig:allspectra}. Only the pn spectra are displayed here for clarity, however the fit was performed jointly on the 3 EPIC detectors. In each case, the color lines highlight the various background components (NXB, CXB, Galactic halo and local hot bubble). The best-fit parameters are given in Table \ref{tab:1}. We note that beyond the tip of the substructure the surrounding ICM emission is significantly below the observed emission (by at least an order of magnitude) and the background components; adding such a component in the spectra does not affect the best-fit parameters provided here.

As shown in Table \ref{tab:1}, the temperature along the structure is roughly constant at temperatures beyond 1.3 and 1.5 keV. Such temperatures are very low compared to the temperature of the cluster \citep[in the range 7-12 keV out 1 Mpc, see Fig. 7 of][]{rossetti13}. Even when considering the temperature decline towards the outer regions, the temperature of the ICM at $R_{500}$ should be approximately 70\% of the average cluster temperature \citep{lm08,pratt07}, which in the case of A2142 corresponds to roughly 6 keV. To confirm this statement, we also extracted the spectrum in a broad region outside the structure and at the same projected radius as the tip to estimate the temperature and density of the local ICM. This region was defined as a segment of annulus centered on the main cluster at projected radii of 12-16 arcmin, i.e. at the same cluster-centric distance as the inner part of the feature. The spectrum of this region is shown in the bottom-right panel of Fig. \ref{fig:allspectra}. The temperature of the undisturbed ICM at this distance from the cluster center is found to be of $6.8_{-1.6}^{+2.1}$ keV, in agreement with the above estimate and significantly hotter than the temperature of the structure. Beyond the tip the ambient cluster emission becomes too weak to enable a reliable measurement of the cluster temperature. 

\begin{table*}
\caption{\label{tab:1}Best-fit parameters for the various regions defined in Fig. \ref{fig:regions} and for the local ICM around the tip of the emission. See Fig. \ref{fig:allspectra} for the best fit and the various spectral components.}
\begin{center}
\begin{tabular}{cccccc}
\hline
Region & Distance [arcmin] & $kT$ [keV] & $K$ arcmin$^{-2}$ & $Z_{\rm Fe}$ & C-stat/d.o.f.\\
\hline
\hline
Tip & $0-1$ & $1.55_{-0.26}^{+0.19}$ & $4.1_{-0.7}^{+0.8}\times10^{-5}$ & $0.19_{-0.09}^{+0.14}$ & 46.43/43\\
Sector 1 & $1-2.2$ &  $1.34_{-0.10}^{+0.22}$ & $2.1_{-0.7}^{+0.3}\times10^{-5}$ & $0.09_{-0.04}^{+0.15}$ & 137.11/127\\
Sector 2 & $2.2-3.5$ & $1.55_{-0.11}^{+0.09}$ & $2.0_{-0.2}^{+0.2}\times10^{-5}$ & $0.26_{-0.07}^{+0.09}$ & 196.7/184\\
Sector 3 & $3.5-5.1$ & $1.30_{-0.09}^{+0.12}$ & $2.3_{-0.2}^{+0.1}\times10^{-5}$ & $0.04_{-0.01}^{+0.02}$ & 326.9/343\\
Sector 4 & $5.1-7.2$ & $1.38_{-0.09}^{+0.26}$ & $7.3_{-1.1}^{+0.9}\times10^{-6}$ & $0.07_{-0.04}^{+0.08}$ & 178.53/182\\
ICM & $12-16$ & $6.8_{-1.6}^{+2.1}$ & $8.1_{-0.7}^{+0.5}\times10^{-6}$ &  & 226.8/192\\
 \hline
\end{tabular}
\end{center}
\textbf{Column description:} 1: Region as defined in Fig. \ref{fig:regions}. 2: Distance from the tip in arcmin. In the case of the ICM region, the values quote the projected distance to the cluster center. 3: Best-fit temperature in keV. 4: Normalization of the APEC model per arcmin$^2$. 5: Metal abundance. 6: Minimum C-statistic and number of degrees of freedom.
\end{table*}

Given the uncertainty in the geometry of the substructure, to compare the thermodynamical properties of the gas inside and outside the substructure we computed the pseudo-entropy ratio $\sigma$ \citep[defined as the ratio between the projected entropies $T_{\rm IN}/T_{\rm OUT}\times(EM_{\rm IN}/EM_{\rm OUT})^{-1/3}$,][]{rossetti10} between the tip and the surrounding ICM. This parameter was found to be tightly correlated with the 3-dimensional entropy ratio. At the tip, we measure $\sigma=0.11\pm0.03$, thus the entropy in the structure appears to be an order of magnitude lower than in the surrounding medium.

The emission measure of the structure is flat between sectors 1 through 3, then drops significantly in sector 4, in agreement with our analysis of the surface brightness (see Fig. \ref{fig:sbtail}). Interestingly, we note that the metallicity of the gas appears to drop significantly between sectors 2 and 3. This can be clearly seen in the spectra of the two regions (see the middle panels of Fig. \ref{fig:allspectra}): while in sector 2 a prominent Fe L-shell blend is observed around 1 keV, corresponding to an iron abundance of roughly 0.3$Z_\odot$, this feature is absent in sector 3, which results in a tightly-constrained metallicity of 0.04 Solar. We note that the emission measure of the plasma is very similar in these two regions, thus this result is unlikely to be caused by the usual degeneracy between emission measure and normalization at low X-ray-emitting temperatures.

\subsection{Dynamics of the NE region}
\label{sec:dynamics}

\begin{figure}
\centerline{\resizebox{\hsize}{!}{\includegraphics{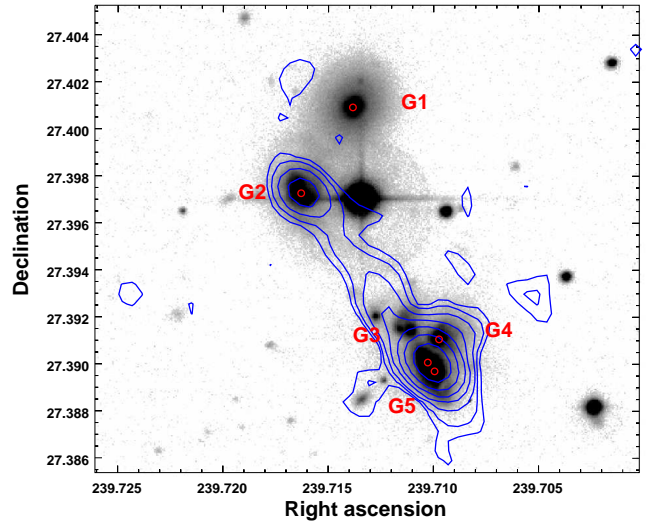}}}
\caption{Same as Fig. \ref{fig:zoom} (right), zoom on the tip of the X-ray emission. The red circles indicate the position of the spectra from the study of \citet{owers11}. The blue contours show the 610 MHz radio emission observed by GMRT (see Sect. \ref{sec:radio}).}
\label{fig:tip}
\end{figure}

\begin{figure}
\centerline{\resizebox{\hsize}{!}{\includegraphics{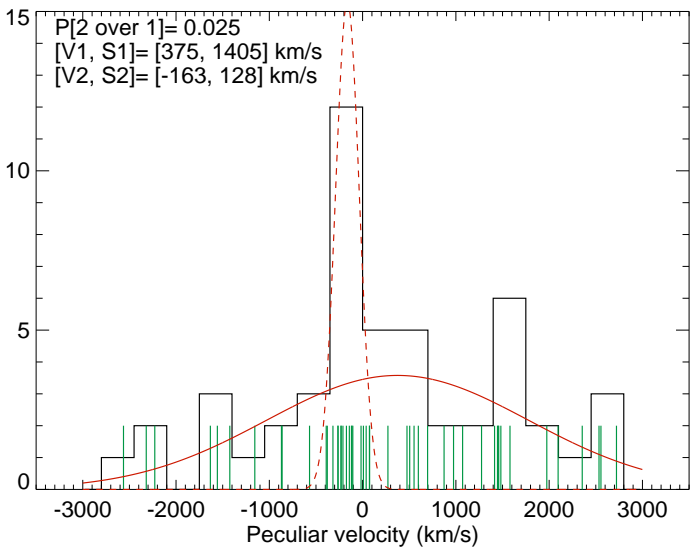}}}
\caption{\label{fig:vpec}The histogram shows the peculiar velocity distribution of the 50 galaxies within $|v_{\rm pec}| < 3500\,$km s$^{-1}$ and $\sim 550$\,kpc of the tip of the NE X-ray feature. The green vertical lines show the $v_{\rm pec}$ positions of each of the 50 galaxies. The solid red and dashed red lines show the main and second Gaussian components from the KMM fit described in Sect. \ref{sec:dynamics}. The best fitting mean and dispersions are listed at the top left of the plot, along with the P-value giving the probability that the observed LRTS of the bimodal to the unimodal fit is due to random fluctuations in a unimodal distribution.}
\end{figure}

A zoom on the CFHT $g$-band image around the tip of the X-ray emission is shown in Fig. \ref{fig:tip}. We identify at least 5 galaxies in this region (labelled G1 through G5) which were spectroscopically classified as cluster members \citep{owers11}. We investigated the dynamics of the region surrounding the tip using the data presented in \citet{owers11} to search for possible substructure in the optical data. There is a mild excess in the local galaxy surface density just to the northwest of the X-ray feature \citep[see Figures 7 and 9 of ][]{owers11}. The peculiar velocity distribution of the 50 galaxies with $|v_{\rm pec}| < 3500$\,km s$^{-1}$ and within a $550\,$kpc radius centered on the X-ray tip is shown in Fig.~\ref{fig:vpec}. 

We searched the distribution for signs of kinematical structure following a similar method to that outlined in \citet{owers11}. Briefly, we use the Kaye's Mixture Modeling algorithm of \citet{ashman94} to fit the distribution with a bimodal Gaussian distribution. For the main component, we use the global cluster dispersion of 995\,km s$^{-1}$ and mean of 0 km s$^{-1}$ as initial estimates to be inputted into the algorithm. Guided by the peak in the distribution at $\sim -200$\,km s$^{-1}$ and the bright elliptical galaxy G1 near the X-ray tip with $v_{\rm pec} = -231\,$km s$^{-1}$, we use initial estimates of the mean, standard deviation and fraction of galaxies to assign of -200\,km s$^{-1}$, 300\,km s$^{-1}$ and $0.2$, respectively, for the second Gaussian component. This resulted in best fits of (mean, sigma) = (375, 1405)\,km s$^{-1}$ for the main cluster and (mean, sigma) = (-163, 128)\,km s$^{-1}$ for the second component. There are 14 galaxies allocated to the second component and the centroid of their spatial distribution is located at (x,y) = (-581, 963)\,kpc with respect to the cluster center. This centroid is $\sim 150\,$kpc SW of the tip of the X-ray emission. To determine if the bimodal fit is statistically preferred over a unimodal fit, we use the parametric bootstrapping method described in \citet{owers12}. Briefly, this is achieved by resampling the best fitting single Gaussian model 5000 times, refitting for both the unimodal and bimodal cases using the same input estimates listed above, and determining the distribution of the likelihood ratio test statistics (LRTS). A LRTS larger than that found for the observed data occurs in $2.5\%$ of the 5000 simulations. Thus, there is marginal ($2\sigma$) evidence favoring the bimodal fit over a unimodal one.

We note that based on its peculiar velocity the compact G3-G5 group shown in Fig. \ref{fig:tip} is likely seen in projection and is probably not part of the substructure discussed here. The peculiar velocities of these galaxies are respectively 970 km~s$^{-1}$ (G3), 1460 km~s$^{-1}$ (G4) and 700 km~s$^{-1}$ (G5), i.e. $>3\sigma$ away from the velocity of the G1 group. The spectra of these galaxies show prominent emission lines with indication of broad components (Owers et al. in prep.). G3 and G5 have line ratios ($O_{\rm III}/H_{\beta}$ and $N_{\rm II}/H_{\alpha}$; as measured from the narrow component of the lines) which place them close to the line dividing AGN and star-forming galaxies on the BPT diagram \citep{baldwin81}, indicating a composite AGN/star-forming source for the ionizing radiation powering the emission. Ongoing star-formation is supported by the stellar continuum which is dominated by Balmer absorption at bluer wavelengths. The line ratios for G4 indicate the dominant ionizing source is an AGN.

\subsection{Radio properties of the NE region}
\label{sec:radio}

We also examined the radio emission in the region around the NE X-ray feature to search for possible effects of the infall on any relativistic plasmas. A2142 is currently being observed with multiple radio telescopes over a wide range of frequencies to study the overall cluster dynamics and their relation to the giant radio halo \citep{farnsworth13}. We report on the smaller-scale 610~MHz emission from the Giant Metrewave Radio Telescope (GMRT) around the  infalling region, for a total of 10 hours of observations on March 23, 2013. Using a a total bandwidth of 32~MHz,standard RFI removal, bandpass correction, a-priori and then phase self-calibration, we produce an image with an angular resolution of 
$5.1^{\prime\prime}\times4.5^{\prime\prime}$ and a noise level $1\sigma\sim 35\mu$Jy~b$^{-1}$.  This is shown in Fig. \ref{fig:tip}, overlaying the CFHT/MegaCam $g$-band image around the tip.

The tip of the X-ray substructure shows a remarkable level of radio activity from the cluster galaxies, although, as discussed below, there is likely more than one kinematical substructure present. Radio emission is clearly associated with at least G2, G4 and G5. The association of the jet- or tail-like feature visible north of G4+G5 is more uncertain; it could originate from the high velocity group galaxy G3 or the fainter object just north-east of G3. The flux density of the blended tip, i.e. G4, G5 and the uncertain G3 identification, is 29.1~mJy, while S$_{\rm 610 MHz}$(G2)=3.0 mJy. Comparison with the radio data available at other frequencies shows that logP$_{\rm 1.6 GHz}$(G2)= 22.5~W/Hz. Considering the difficulty in separating the emission of the individual galaxies in the tip, it is reasonable to assume that their radio powers are at a similar level to G2. 

The fraction of radio-loud galaxies in these infalling structures is unexpectedly high.  At the radio luminosity of G2 (and likely G4, G5 and possibly G3), only 2\% to 15\% of optical galaxies are typically detected \citep[e.g.,][]{ledlow96,mauch07}.  It is not yet clear whether this high level of activity is directly related to the infall process and the stripping of the X-ray plasma. Existing studies attempting to connect the dynamical state of cluster galaxies or substructures to radio AGN or starburst activity are so far inconclusive \citep[e.g.][]{miller03,venturi00}.  

\section{Discussion}
\label{sec:disc}

\subsection{Interpretation of the NE substructure}

Our analysis of the \emph{XMM-Newton} indicates that the gas in the substructure is significantly cooler ($kT=1.3-1.5$ keV) than its environment. At the tip of the substructure, the (pseudo-)entropy of the plasma is an order of magnitude lower than in the local ICM; this implies that this gas cannot be associated with the virialized cluster ICM. Therefore, this feature is caused by a secondary halo. The observed average temperature of 1.4 keV in the substructure is typical of the virialized plasma of a galaxy group with a mass of a few $10^{13}M_\odot$ \citep{sun09}. As shown in Fig. \ref{fig:zoom}, the tip of the X-ray substructure corresponds with an overdensity of cluster galaxies. Our dynamical analysis of the galaxies in this region (see Sect. \ref{sec:dynamics}) shows the presence of a mildly-significant ($2\sigma$) substructure in this region at $v_{\rm pec}\sim-200$ km s$^{-1}$. Moreover, the peak of the X-ray emission coincides with the bright early-type galaxy G1 (see Fig. \ref{fig:tip}) at $v_{\rm pec}=-231$ km s$^{-1}$. Therefore, although the presence of this substructure cannot be firmly established based on the optical data, we interpret G1 as the dominant galaxy of the substructure and we associate it with the X-ray feature. 

Given all this information, we identify the NE substructure discovered here with an infalling galaxy group in the process of being disrupted by the interaction with the main cluster. In this paradigm, the tip of the X-ray emission corresponds with the remaining unperturbed gas lying at the center of the group. The bulk of the X-ray emission, which lags behind the position of the group, comes from intragroup gas which has been stripped from its original dark-matter halo by the ram pressure of the ambient ICM. The low peculiar velocity of the group indicates that it is moving mostly along the plane of the sky in the direction of the cluster center. In this case, we predict the existence of a cold front at the tip pointing in the SW direction. Unfortunately, the moderate angular resolution of \emph{XMM-Newton} and the presence of a point source just ahead of the tip prevent us from detecting such a feature; observations of this region with \emph{Chandra} are required to test this scenario.

\subsection{Ram-pressure stripping properties}

With a projected scale of at least 800 kpc (see Fig. \ref{fig:sbtail}), this is the longest X-ray tail reported to date. It is at least twice as long as the X-ray tail of the massive elliptical M86 in the Virgo cluster \citep{randall08}, which exhibits a projected size of 150 kpc and a reconstructed 3D length of 380 kpc. Unlike most of the X-ray tails reported thus far, the bulk of the X-ray emission is located in the tail rather than in the group; this suggests that the group is already in an advanced state of disruption, in spite of its large distance from the cluster core. 

A simple estimate of the velocity of the moving group can be obtained under the hypothesis that the gas at the tip of the structure is in pressure equilibrium with its surrounding ICM \citep{mark00},

\begin{equation} P_{\rm ICM}  + \rho v_{\rm rel}^2 \approx P_{\rm group} . \label{eq:ram}\end{equation}

\noindent To estimate the gas density and pressure at the tip of the structure, we assume that the emission originates from a spherical region of radius $0.5$ arcmin and that the density is constant within this region. Using these assumptions we estimate an average electron density of $(1.8\pm0.2)\times10^{-3}$ cm$^{-3}$ at the tip. To compare with the gas density in the ambient plasma, we used the radially-averaged density profile of the cluster extracted by \citet{e12} using \emph{ROSAT}/PSPC data, which extends out to 2.5 Mpc from the cluster core. The mean density of the ICM at 1.2 Mpc from the cluster center is $(1.6\pm0.2)\times10^{-4}$ cm$^{-3}$. Given the difference in temperature between the regions inside and outside the tip (see Table \ref{tab:1}), we indeed find that the tip is over-pressured compared to the surrounding medium: $P_{\rm group}=(4.4\pm0.8)\times10^{-12}$ ergs cm$^{-3}$, $P_{\rm ICM}=(1.7\pm0.5)\times10^{-12}$ ergs cm$^{-3}$, which could be explained by a significant ram pressure (Eq. \ref{eq:ram}). This translates into a value $v_{\rm rel}=1,260\pm240$ km s$^{-1}$ for the relative velocity of the group. Such a velocity is typical for this kind of systems and comparable to the adiabatic sound speed in the medium ($c_s\sim1100$ km s$^{-1}$ in a 7 keV plasma). Assuming that the group has been moving at constant velocity, this calculation implies that the gas located in the outermost region of the structure ($\geq800$ kpc from the tip) was expelled from the group $\geq600$ Myr ago. 

To estimate the gas mass within the tail, we approximated the geometry of each region defined in Fig. \ref{fig:regions} as a segment of cone \citep[which is the geometry expected for ram-pressure stripping of a spherical hot halo,][]{toniazzo01} and assumed constant density into each segment\footnote{It must be noted that the numbers provided here are rough estimates given the uncertainty in the geometry of the tail}. Based on the emission measures reported in Table \ref{tab:1} we compute the average density and the gas mass in each region along the tail. Using these assumptions we estimate that the gas density progressively decreases along the tail from $\sim2\times10^{-3}$ at the tip down to $\sim3\times10^{-4}$ at the NE end. Integrating over the volume, we obtain a gas mass of $\sim1.4\times10^{12}M_\odot$ in the tail, which is considerable. For comparison, this is more than 2 orders of magnitude larger than the gas mass in the tail of M86 \citep{randall08}. Given the approximate age of the feature ($\geq600$ Myr, see above), we infer a substantial mass-loss rate of $\sim2,300M_\odot$ yr$^{-1}$. For a group of mass $3\times10^{13}M_\odot$ \citep{sun09}, the observed gas mass corresponds to roughly 5\% of the total mass of the group, which is close to the hot gas fraction usually reported for such groups \citep{gasta07,sun09}. We are thus observing a large fraction of the original intragroup plasma being stripped away.

To compare these numbers with the expectations of simple ram-pressure stripping estimates \citep{gunn72,nulsen82}, we assumed that prior to the infall the pressure profile of the group was described by a universal pressure profile \citep{arnaud10}. In this paradigm, all the intragroup medium with a thermal pressure $P_{th}<P_{ram}$ (see Eq. \ref{eq:ram}) can be stripped in a short timescale \citep{mori00}. Using the universal pressure profile, we find that the ram pressure exceeds the thermal pressure of the group beyond $\sim0.1 R_{200, {\rm  group}}$ and the gas beyond this radius can be stripped rapidly, such that the subcluster may have already lost $>90\%$ of its total gas mass. Using the above assumptions on the geometry of the system, we estimate that the gas mass remaining within the tip is $\sim5\times10^{10}M_\odot$, i.e. $\sim5\%$ of the gas mass of the tail. Therefore, we conclude that our observations are consistent with the expectations of ram-pressure stripping. Our results demonstrate that during infall the hot gas content of such groups is efficiently stripped already at such large distances from the cluster core, in agreement with the recent simulations of \citet{cen14}.

\subsection{Thermal conduction in the local ICM}

The very large extent of the stripped tail allows us to set constraints on the effective thermal conductivity in the ICM of A2142. Indeed, thermal conduction quickly transfers heat between the hot and cold phase. This has the consequence of washing out not only temperature gradients, but also density variations, thereby smoothing the observed X-ray surface brightness \citep{gaspari13}. The conductivity timescale in a plasma can be estimated as

\begin{equation} t_{\rm cond} \approx \frac{\ell^2}{D_{\rm cond}} = \frac{3n_e \ell^2 k_B}{2\kappa},  \end{equation}

\noindent where $D_{\rm cond}$ is the diffusivity of thermal conduction, $n_e$ is the electron number density in the surrounding medium, $\ell$ is the length scale of the region, $k_B$ is the Boltzmann constant, and $\kappa$ is the conductivity coefficient in the medium. The latter can be written as a fraction of the \citet{spitzer} plasma conductivity, 

\begin{equation} \kappa=f\kappa_S=f1.84\times10^{-5}T^{5/2}(\ln{\Lambda})^{-1},\end{equation}

\noindent where $\Lambda$ is the Coulomb logarithm \citep[see][]{gaspari13} and $f$ is the effective isotropic suppression factor, incorporating magnetic fields effects on small scales. The temperature of the cluster at $R_{200}$ is $\sim3.5$ keV \citep{aka11}; since the outermost region of the tail is located halfway between $R_{500}$ and $R_{200}$, we adopt a value of 5 keV for the temperature of the gas. In this region the \emph{ROSAT} data indicate a density of $\sim5\times10^{-5}$ cm$^{-3}$ \citep{e12}, thus $\kappa_S\sim1.2\times10^{13}$ ergs s$^{-1}$ cm$^{-1}$ K$^{-1}$. The relevant length scale is $\ell\sim75$ kpc as the maximum distance from the ICM (half width) to evaporate the structure at its NE end. Therefore, we retrieve a conduction timescale $t_{\rm cond} = 1.4f^{-1}$ Myr.

As a result, the Spitzer conductivity timescale in the outermost region of the tail is a factor $\gtrsim400$ shorter than the survival timescale of this plasma ($\geq600$ Myr, see above). The discovery of the infalling group in A2142 thus provides evidence that thermal conduction is largely suppressed (by a factor $f^{-1}\gtrsim400$) in the cluster's environment. Similar conclusions have been reached in the past few years for special environments such as cold fronts \citep{ettori00} and filamentary structures in cluster cores \citep{forman07,sanders14}, and from the power spectrum of density fluctuations \citep{gaspari13,gaspari14}. This is however the first time that this result is confirmed in a direct way outside the cluster core.

In the few magneto-hydrodynamics (MHD) simulations including turbulence and anisotropic conduction, the retrieved suppression factor is of the order of $\sim3-5$ \citep[see also \citealt{narayan01}]{ruszkowski10}. This arises from the simple geometric suppression in a chaotic field, since electrons can diffuse only along the B-field lines. However, if the B-field coherence length is smaller than the plasma mean free path $\lambda_e$, the spatial divergence of neighboring field lines in combination with the trapping of electrons between magnetic mirrors can induce suppression factors up to  $10^2-10^3$ \citep{chandran98,rechester78}. Large suppression factors thus suggest a small coherence length ($<$ kpc), which may be a simple by-product of highly chaotic turbulent motions in the ICM. On top of that, plasma microinstabilities \citep[e.g. firehose,][]{scheko10} acting on scales $\ll \lambda_e$ may drastically change the transport properties, leading to further suppression \citep{mcnamara12}. Simulating the kpc region and the gyroradius scales (for a $\mu$G field in a $10^8$ K plasma, the gyroradius is $10^{-10}$ pc!) in the same kinetic simulation is currently beyond the reach of numerical simulations.

It has been also suggested \citep{asai05} that subclusters moving within the atmosphere of larger systems develop a magnetic drape inhibiting transport processes, but this scenario seems unlikely to inhibit conduction by more than a factor of a few \citep{zuhone13}. Moreover, \citet{ruszkowski14} found that magnetic draping does not suppress the rate of ram-pressure stripping; the B-field can actually enhance it. The magnetic field becomes highly tangled near the transition layer due to gravitational instabilities and velocity gradients. In combination with the build-up of magnetic pressure, this undoes the protective effect of draping, allowing the gas and heat to substantially leak out. The breaking of the drape implies that other processes (as described above) are required to suppress conduction by many orders of magnitude.

We note that this result should also have implications on the fate of the stripped gas. Indeed, tangling of the magnetic field should also affect other transport processes, thus viscosity will likely be suppressed by a factor which is similar to the one affecting conductivity\footnote{Note however that viscosity is tied with ions, while heat is transported by electrons}. If the infalling plasma is stable against Kelvin-Helmholtz instabilities \citep{roediger13}, the low-entropy plasma present within the the group would, in a few dynamical timescales, settle within the core of the cluster, where ICM with similar entropy is found. Thus, infalling substructures would  ``donate" a significant fraction of their cool gas to the cluster core, possibly forming filamentary structures such as the ones observed in the core of Coma \citep{sanders14}. In the future, high-resolution imaging X-ray spectrometers such as those that will equip \emph{ASTRO-H} \citep{astroh} and \emph{Athena} \citep{athenawp} will allow us to test this scenario and probe directly the physics of the ICM plasma through the study of bright X-ray emission lines.

\subsection{Metallicity along the substructure}

As shown in Table \ref{tab:1} and discussed in Sect. \ref{sec:spectra}, the metallicity of the gas appears to drop from the tip to the outermost regions of the tail. In particular, an abrupt change of metallicity can be noticed between sectors 2 and 3 (see Fig. \ref{fig:regions}). This drop could be indicative of the original metallicity gradient of the group. Indeed, as the group gradually enters the cluster's atmosphere the gas density increases, and thus the ram pressure $\rho v_{\rm rel}^2$ becomes stronger. Therefore, during infall gas with progressively higher thermal pressure gets stripped. In other terms, the outskirts of the group can be stripped first and will now be localized in the outermost region of the tail, whereas the regions close to the core of the group must have been stripped recently and lie in the innermost parts of the tail. 

In a sample of 15 nearby groups, \citet{rasmussen09} found that the typical metallicity of the hot gas in galaxy groups decreases from near Solar values in the core to low metallicities ($\sim0.1Z_\odot$) in the outskirts. Therefore, our results on the metallicity appear to confirm the above picture that the gas at larger distance from the tip comes from larger radii in the original dark-matter halo of the group. Deeper X-ray observations could allow us to trace the original radial profile of the group through a more detailed analysis of the metal distribution.

We note however that the average mass-weighted metallicity of the gas is $\sim0.12$, which is low compared to the typical metallicity of clusters and groups \citep[$\sim0.25$,][]{lm08b}. In particular, the metallicity observed in Sector 3 ($0.04Z_\odot$) is very low compared to the expected values. As a possible explanation, we considered the possibility that the metal abundance measured in the outermost regions of the tail might be biased low by a significant multiphase structure of the intragroup medium. In case the material of this region is in the process of mixing with its surroundings, the plasma might be clumpy and multiphase, which could bias the measured metallicity when assuming a single-temperature plasma. To test this hypothesis, we fitted the data of Sector 3 with the differential emission-measure plasma model \texttt{c6mekl} \citep{singh96}, fixing the metal abundance to the canonical value of $0.3Z_\odot$. This model returns an equally-good fit to the data as the single-temperature model, thus the plasma in these regions might indeed be significantly multiphase. Higher spectral resolution data are required to understand whether the metallicity is genuinely low in this region or if this result is caused by a multiphase emission-measure distribution.

\section{Conclusions}

In this paper, we have reported the discovery of a diffuse, irregular X-ray feature in the outskirts of the massive cluster Abell 2142, which we associate with an infalling galaxy group. Our findings can be summarized as follows:

\begin{itemize}
\item The tip of the X-ray substructure coincides with an overdensity of cluster galaxies with a peculiar velocity of $-200$ km s$^{-1}$ with respect to the mean cluster redshift. Thus, we identify this structure as the core of the infalling group. 
\item The tip is the brightest region of the X-ray substructure and points towards the cluster center, which indicates that we are witnessing the infall of the group in the potential well of the main cluster. This interpretation is further supported by our measurement of the gas temperature of the structure ($\sim1.4$ keV), which is significantly lower than the ambient ICM ($\sim7$ keV) and typical of the virialized plasma of a galaxy group with a mass of a few $10^{13}M_\odot$. The entropy of the gas in the substructure is an order of magnitude lower than in the ICM at the same radius, which confirms that the gas cannot be associated with the main cluster.
\item The bulk of the X-ray emission is trailing behind the galaxy group over a linear scale of about 800 kpc (8 arcmin). We interpret this feature as gas originally belonging to the galaxy group which has been stripped from its original dark-matter halo by the ram pressure of the ambient ICM. This is the longest X-ray tail reported to date. Assuming a cone geometry, the expelled gas mass is of the order of $\sim1.5\times10^{12}M_\odot$, which corresponds to a large fraction of the original intragroup medium \citep{gasta07}.
\item Under the hypothesis that the gas at the tip is in pressure equilibrium with its surrounding medium, we estimate the relative velocity of the subcluster to be $\sim1,200$ km s$^{-1}$. Assuming that the thermal pressure distribution of the group was originally described by a universal pressure profile \citep{arnaud10}, we estimate that even at such large distance from the core the cluster's ram pressure is sufficient to expel $>90\%$ of the gas mass of the group. This is confirmed by the gas remaining within the tip, which amounts to just 5\% of the gas mass of the tail.
\item Given the observed length of the structure ($\geq800$ kpc), we infer that the warm intragroup gas has been surviving in the local ICM for at least 600 Myr. This exceeds the Spitzer conductivity timescale in the surrounding medium by a factor of $\geq400$. We interpret this result as evidence for a tangled magnetic field configuration in the ICM with a coherence length smaller than the electron mean free path \citep{chandran98}.
\item The long survival of the intragroup plasma in contact with the surrounding ICM implies that the low-entropy gas within the tail may eventually settle within the core of the cluster, where gas with similar entropy resides. This would have strong implications on the evolution of the hot gas content of galaxy clusters.
\item The metal abundance of the gas decreases from the tip to the outermost regions of the tail. We explain this result as evidence that the gas in the most distant regions of the tail was originally located at large radii in the group halo and was stripped first because of its low thermal pressure.
\end{itemize}

\acknowledgements Based on observations obtained with \emph{XMM-Newton}, an ESA science mission with instruments and contributions directly funded by ESA Member States and NASA. We thank the staff of the GMRT that made the radio observations possible. GMRT is run by the National Centre for Radio Astrophysics of the Tata Institute of Fundamental Research. DE thanks Elke Roediger, Jean-Paul Kneib and HuanYuan Shan for useful discussions. Partial support for LR comes from US NSF Grant AST-1211595 to the University of Minnesota.

\bibliographystyle{aa}
\bibliography{group}

\end{document}